\documentclass[sigconf]{acmart}
\pdfoutput=1
    
\copyrightyear{2023}
\acmYear{2023}
\setcopyright{rightsretained}
\acmConference[WWW '24]{The Web Conference}{May 13--17, 2024}{Singapore}

\usepackage[inline]{enumitem}

\copyrightyear{2024}
\acmYear{2024}
\setcopyright{rightsretained}
\acmConference[WWW '24 Companion]{Companion Proceedings of the ACM Web Conference 2024}{May 13--17, 2024}{Singapore, Singapore}
\acmBooktitle{Companion Proceedings of the ACM Web Conference 2024 (WWW '24 Companion), May 13--17, 2024, Singapore, Singapore}\acmDOI{10.1145/3589335.3641303}
\acmISBN{979-8-4007-0172-6/24/05}

\begin{document}

\title{The 2nd Workshop on Recommendation with Generative Models}
\subtitle{Website: \url{https://generative-rec.github.io/workshop/}}

\author{Wenjie Wang}
\email{wangwenjie@u.nus.edu}
\affiliation{%
  \institution{National University of Singapore}
  \country{}
}

\author{Yang Zhang}
\email{zy2015@mail.ustc.edu.cn}
\affiliation{
\institution{University of Science and Technology of China}
\country{}
}
\author{Xinyu Lin}
\email{xylin1028@gmail.com}
\affiliation{%
  \institution{National University of Singapore}
  \country{}
}

\author{Fuli Feng}
\email{fulifeng93@gmail.com}
\affiliation{%
  \institution{University of Science and Technology of China}
  \country{}
}

\author{Weiwen Liu}
\email{liuweiwen8@huawei.com}
\affiliation{
  \institution{Huawei Noah's Ark Lab}
  \country{}
}
\author{Yong Liu}
\email{liu.yong6@huawei.com}
\affiliation{%
  \institution{Huawei Noah's Ark Lab}
  \country{}
}
\author{Xiangyu Zhao}
\email{xianzhao@cityu.edu.hk}
\affiliation{%
  \institution{City University of Hong Kong}
  \country{}
}
\author{Wayne Xin Zhao}
\email{batmanfly@gmail.com}
\affiliation{%
  \institution{Renmin University of China}
  \country{}
}

\author{Yang Song}
\email{yangsong@kuaishou.com}
\affiliation{%
  \institution{Kuaishou Technology}
  \country{}
}

\author{Xiangnan He}
\email{xiangnanhe@gmail.com}
\affiliation{%
  \institution{University of Science and Technology of China}
  \country{}
}


\renewcommand{\shortauthors}{Wenjie Wang et al.}

\begin{abstract}

The rise of generative models has driven significant advancements in recommender systems, leaving unique opportunities for enhancing users' personalized recommendations. This workshop serves as a platform for researchers to explore and exchange innovative concepts related to the integration of generative models into recommender systems. 
It primarily focuses on five key perspectives: 
\begin{enumerate*}[label=(\roman*)]
\item improving recommender algorithms, 
\item generating personalized content,  
\item evolving the user-system interaction paradigm, 
\item enhancing trustworthiness checks, and 
\item refining evaluation methodologies for generative recommendations.  
\end{enumerate*}
With generative models advancing rapidly, an increasing body of research is emerging in these domains, underscoring the timeliness and critical importance of this workshop. 
The related research will introduce innovative technologies to recommender systems and contribute to fresh challenges in both academia and industry. 
In the long term, this research direction has the potential to revolutionize the traditional recommender paradigms and foster the development of next-generation recommender systems. 

\vspace{-0.3cm}
\end{abstract}

\ccsdesc[500]{Information systems~Recommender systems}
\keywords{Generative Models for Recommendation, Large Language Models, Trustworthy Recommendation}

\maketitle

\vspace{-0.3cm}
\section{Scope and topics}
\label{scope}

The main objective of this workshop is to encourage pioneering research in the integration of generative models with recommender systems, with a specific focus on five key aspects. 
First, this workshop will motivate active researchers to utilize generative models for enhancing recommender algorithms and refining user modeling. 
Second, it promotes utilizing generative models to generate diverse content, \textit{i.e.,} AI-generated content (AIGC), in certain situations, complementing human-generated content to satisfy a broader range of user preferences and information needs. 
Third, it embraces substantial innovations in user interactions with recommender systems, possibly driven by the boom of large language models (LLMs). 
Fourth, the workshop will highlight the significance of trust in employing generative models for recommendations, encompassing aspects like content trustworthiness, algorithmic biases, and adherence to evolving ethical and legal standards. 
Lastly, the workshop will prompt researchers to develop diverse methods for the evaluation, including novel metrics and human evaluation approaches. 

The workshop provides an invaluable forum for researchers to present the latest advancements in the rapidly evolving field of recommender systems. We welcome original submissions focusing on generative models in recommender systems, including a range of relevant topics: 
\begin{itemize}[leftmargin=*]
    
    \item Leveraging LLMs and other generative models such as diffusion models to improve user modeling and various recommendation tasks, including sequential, cold-start, social, conversational, multimodal, and causal recommendation tasks. 
    \item Improving generative recommender models (\textit{e.g.,} LLM-based recommenders) from different aspects, such as model architecture, and training and inference efficiency. 
    \item Combining external knowledge from LLMs or other generative models to enhance user and item representation learning. 
    \item Generative recommendation by harnessing generative AI to drive personalized item creation or editing, particularly in contexts such as advertisement, image, and micro-video. 
    \item Innovation of user-system interaction paradigm for effective user feedback by leveraging strong conversational capability of LLMs. 
    \item Real-world applications of generative recommender systems, ranging from finance to streaming platforms and social networks. 
    \item Trustworthy recommendation with generative models, for example, developing the standards and technologies to improve or inspect the recommendations from the aspects of bias, fairness, privacy, safety, authenticity, legal compliance, and identifiability. 
    \item Developing generative agents empowered by LLMs, motivating the recommendation agents from user simulation and data collection, to algorithm enhancement and evaluation. 
    \item Evaluation of generative recommender systems, including new evaluation metrics, standards, and human evaluation approaches.
\end{itemize}

\section{Rationale}

\subsection{Relevance}

This workshop, centered on recommendation using generative models, aligns seamlessly with the Web conference, as it spotlights a pivotal research trend within recommender systems, which is an essential area at the Web conference. 
Moreover, incorporating generative models into recommender systems can improve information filtering services to cater to users' personalized information needs in a variety of Web applications. 

\subsection{Objectives and Expected Outcome}

This workshop will encourage researchers to venture into new horizons within recommender systems by incorporating powerful generative models. 
These ongoing endeavors and emerging technologies will introduce fresh characteristics to industry products and motivate innovative research topics in academia. 
Extensive applications, spanning streaming media, social networks, and forums, are well-positioned to embrace these generative model techniques. 
In the long run, this research direction holds the potential to revolutionize the established recommender paradigm, giving rise to the evolution of next-generation recommender systems. 

We expect to receive innovative contributions in this promising field with exciting ideas and novel methodologies. Some high-quality papers can also be encouraged to be submitted an extension version to our organized ACM TOIS special issue on using pre-trained models for recommendation.

\subsection{Target Audience}

This workshop's attractiveness stems from its dedication to an evolving area of recommender systems~\cite{wang2023diffrec, lin2023multi, wang2023generative, cui2022m6, li2023gpt4rec, liu2023chatgpt, wang2023zero, gao2023chat, Zhang2021,deffayet-2023-generative}. 
It aims to draw the attention of a diverse audience, including researchers, industry experts, and academics. The workshop offers a unique forum for these stakeholders to share innovative ideas, methods, and accomplishments, encouraging interdisciplinary collaboration and the exploration of novel applications.



\subsection{Related Workshops}

We have initiated the first workshop on recommendation with generative models, engaging many researchers and experts in the community to discuss research progress. The first workshop is held at the 32nd ACM International Conference on Information and Knowledge Management (CIKM '23), in Birmingham UK, on October 22nd. It has attracted extensive attention with around 50 offline participants and over 60 online attendees via Zoom. 
The first workshop accepted 9 high-quality contributions out of 11 submissions, covering the research papers and surveys on LLM-based recommendation and diffusion recommender models. 
More information is at \url{https://rgm-cikm23.github.io/}. Besides, the Gen-IR workshop\footnote{\url{https://coda.io/@sigir/gen-ir}.} at SIGIR'23 encourages the exploration of generative information retrieval. However, this workshop is more about general information retrieval, which is not specialized in leveraging generative models to drive the advancements in recommender systems. 

The rapid progress of generative models for recommendation has spurred a wave of innovative research efforts in the past months~\cite{wang2023diffrec,lin2023multi,geng2023vip5,mei2023lightlm,zhang2023generative,xi2023towards}. 
Notably, some recent studies explore new research directions, including improvements in {data/model efficiency and architecture of generative recommendation~\cite{lin2024data,mei2023lightlm,li2023prompt}, multimodal generative recommendation~\cite{geng2023vip5,zhou2023gpt4v}, personalized outfit recommendation~\cite{xu2024difashion}, LLM-based federated recommendation~\cite{zhao2024llm}, and generative agents for recommendation~\cite{zhang2023generative,wang2023recagent,zhang2024prospect}.} 
As such, it is very necessary to host the second workshop to foster discussions on these evolving directions. 

\section{Workshop Program Format}


This workshop will be held for \textbf{half a day}. 
We will invite two researchers in this field to give 30-minute keynote talks. We will also invite several senior researchers and developers to organize a panel discussion on future directions. 
{Besides, in this workshop, we have received 10 submissions and accepted 8 papers as follows}: 
\begin{itemize}[leftmargin=*]\small
    \item Diffusion Recommendation with Implicit Sequence Influence. Yong Niu, Xing Xing, Zhichun Jia, Ruidi Liu, Mindong Xin and Jianfu Cui. 
    \item A Study of Implicit User Unfairness in Large Language Models for Recommendation. Chen Xu, Wenjie Wang, Yuxin Li, Liang Pang, Jun Xu and Tat-Seng Chua. 
    \item Aligning GPTRec with Beyond-Accuracy Goals with Reinforcement Learning. Aleksandr Vladimirovich Petrov and Craig Macdonald. 
    \item Controllable and Transparent Textual Latents for Recommender Systems. Emiliano Penaloza, Haolun Wu, Olivier Gouvert and Laurent Charlin. 
    \item How Reliable is Your Simulator? Analysis on the Limitations of Current LLM-based User Simulators for Conversational Recommendation. Lixi Zhu, Xiaowen Huang and Jitao Sang.
    \item Multimodal Conditioned Diffusion Model for Recommendation. Haokai Ma, Yimeng Yang, Lei Meng, Ruobing Xie and Xiangxu Meng.
    \item Bridging Items and Language: A Transition Paradigm for Large Language Model-Based Recommendation. Xinyu Lin, Wenjie Wang, Yongqi Li, Fuli Feng, See-Kiong Ng and Tat-Seng Chua.
    \item OutfitGPT: LLMs as Fashion Outfit Generator and Recommender. Yujuan Ding, Junrong Liao, Wenqi Fan, Yi Bin and Qing Li.
\end{itemize}
Each paper would be given around 10 minutes for presentation and QA. 
We present the preliminary program schedule\footnote{Kindly note that the paper acceptance plan and the program schedule are tentative and may be subject to potential adjustments according to the requirements of the conference chairs.} as follows: 

\begin{table}[ht]
\centering
\setlength{\abovecaptionskip}{0cm}
\setlength{\belowcaptionskip}{0cm}
\caption{Program schedule.}
\label{tab:program}
\begin{tabular}{ll}
\toprule
Event & Time\\
\midrule
Opening Remarks from Co-Chairs & 08:40--08:50 \\
Keynote Talk \#1 followed by QA & 08:50--09:20 \\
Paper Session \#1 ({4} papers) & 09:20--10:00 \\
Tea Break & 10:00--10:30 \\
Keynote Talk \#2 followed by QA & 10:30--11:00 \\
Paper Session \#2 (4 papers) & 11:00--11:40 \\      
Panel Discussion \& Closing Remarks & 11:40--12:00\\
\bottomrule
\end{tabular}
\vspace{-2pt}
\end{table}

\section{Call for papers}
    \subsection{Introduction}
    The surge in generative models has catalyzed substantial progress within recommender systems. 
    For example, pre-trained generative models have demonstrated their capability to effectively learn user preferences from historical interactions~\cite{bao2023tallrec,cui2022m6}; 
    generative models might help to produce item content to meet users' diverse information needs in some scenarios~\cite{wang2023generative}, 
    generative models have shown promise in generating item content that caters to the diverse users' information needs in specific contexts; 
    and the emergence of ChatGPT-like language models offers novel interaction modes to obtain users' feedback and intension~\cite{liao2023proactive,tang2021high}.

    In this light, user experience can be potentially enhanced by advancing the traditional recommender paradigms via generative models. 
    This workshop provides a platform to facilitate the integration of generative models into recommender systems, with a focus on user modeling, content generation, interaction patterns, trustworthiness evaluations~\cite{zhang2023chatgpt}, and evaluation methods~\cite{Sun-2023-evaluation}.

    \subsection{Objectives and Scope}
    This workshop aims to encourage innovative research on integrating generative models with recommender systems, particularly on five key aspects: 
    \begin{enumerate*}[label=(\roman*)] 
    \item enhancing algorithms for user modeling by generative models; 
    \item generating personalized content to supplement human-generated content; 
    \item evolving the user interaction modes with recommender systems; 
    \item prioritizing trustworthiness in generative recommendation; 
    \item formulating evaluation techniques for generative model-based recommender systems. 
    \end{enumerate*}
    We summarize the detailed objectives and scope in Section~\ref{scope}.

    \subsection{Submission Information}
    
    Submission Guidelines:
    Submitted papers must be a single PDF file in the template of ACM WWW 2024. 
    Submissions can be of varying length from 4 to 8 pages, plus unlimited pages for references. 
    The authors may decide on the appropriate length of the paper as no distinction is made between long and short papers. 
    All submitted papers will follow the "double-blind" review policy and undergo the same review process and duration. 
    Expert peer reviewers in the field will assess all papers based on their relevance to the workshop, scientific novelty, and technical quality. 
The timeline is as follows: 
    \begin{itemize}
        \item Submissions deadline: February 26, 2024 
        \item Paper acceptance notification: March 4, 2024 
        \item Workshop date: May 13, 2024
    \end{itemize}

\section{Organizers}
    \begin{itemize}[leftmargin=*]
        
    \item \smallskip
    \noindent\textbf{Wenjie Wang}\\
    \underline{Email}: wangwenjie@u.nus.edu\\
    \underline{Affiliation}: National University of Singapore\\
    \underline{Biography}: Dr. Wenjie Wang is a research fellow at National University of Singapore (NUS). He received Ph.D. in Computer Science from NUS, supervised by Prof. Tat-Sent Chua. Dr. Wang was a winner of Google Ph.D. Fellowships. His research interests cover recommender systems, data mining, and causal inference. His first-author publications appear in top conferences and journals such as SIGIR, KDD, WWW, WSDM, and TOIS. His work has been selected into ACMMM 2019 Best Paper Final List. 

    \item \textbf{Yang Zhang}\\
    \underline{Email}: zy2015@mail.ustc.edu.cn\\
    \underline{Affiliation}: University of Science and Technology of China\\
    \underline{Biography}: 
    Yang Zhang is a Ph.D. candidate at the University of Science and Technology of China (USTC), under the supervision of Prof. Xiangnan He. His research interests lie in recommender systems and causal inference, and he has published several first-author papers in top conferences. In particular, he received the Best Paper Honorable Mention in SIGIR 2021 for his work on causal recommendation. 

    \item \textbf{Xinyu Lin}\\
    \underline{Email}: xylin1028@gmail.com\\
    \underline{Affiliation}: National University of Singapore\\
    \underline{Biography}:  
    Xinyu Lin is a Ph.D. candidate at the University of Singapore, under the supervision of Prof. Tat-seng Chua. Her research interests lie in recommender systems, and her work has been published in top conferences and journals such as SIGIR, WWW, CIKM, and TOIS. 
    Moreover, she has also served as the reviewer and PC member for the top conferences and journals, including SIGIR, WSDM, and TOIS. 
    
    \item \textbf{Fuli Feng}\\
    \underline{Email}: fulifeng93@gmail.com\\
    \underline{Affiliation}: University of Science and Technology of China\\
    \underline{Biography}: Dr. Fuli Feng is a professor in University of Science and Technology of China. He received Ph.D. in Computer Science from National University of Singapore in 2019. His research interests include information retrieval, data mining, causal inference, and multi-media processing. He has over 60 publications appeared in several top conferences such as SIGIR, WWW, and SIGKDD, and journals including TKDE and TOIS. He has received the Best Paper Honourable Mention of SIGIR 2021 and Best Poster Award of WWW 2018. 
    Moreover, he organized the 1st workshop on Information Retrieval in Finance at SIGIR'20. 

    \item \textbf{Weiwen Liu}\\
    \underline{Email}: liuweiwen8@huawei.com\\
    \underline{Affiliation}: Huawei Noah's Ark Lab\\
    \underline{Biography}: Dr. Weiwen Liu is currently a senior researcher at Huawei Noah's Ark Lab. She received her Ph.D. in Computer Science and Engineering from the Chinese University of Hong Kong in 2020. Her research is broadly concerned with recommender systems, information retrieval, and user preference learning. She has published over 40 papers on top conferences including KDD, SIGIR, and WWW. She gave a tutorial on neural re-ranking at RecSys'22. She will co-organize the DLP workshop at RecSys'23. 
    
    \item \textbf{Yong Liu}\\
    \underline{Email}: liu.yong6@huawei.com\\
    \underline{Affiliation}: Huawei Noah's Ark Lab\\
    \underline{Biography}: Dr. Yong Liu is a Senior Principal Researcher at Huawei Noah's Ark Lab, Singapore. Prior to joining Huawei, he was a Senior Research Scientist at Nanyang Technological University (NTU), a Data Scientist at NTUC Enterprise, and a Research Scientist at Institute for Infocomm Research (I2R), A*STAR, Singapore. 
    Moreover, he has served as the Challenge Co-chair for RecSys 2023, PC Co-chair for ICCSE 2021. 

    \item \textbf{Xiangyu Zhao}\\
    \underline{Email}: xianzhao@cityu.edu.hk\\
    \underline{Affiliation}: City University of Hong Kong\\
    \underline{Biography}:
    Prof. Xiangyu Zhao is a tenure-track assistant professor of Data Science at City University of Hong Kong (CityU). His research has been awarded ICDM'22 and ICDM'21 Best-ranked Papers, Global Top 100 Chinese New Stars in AI, CCF-Tencent Open Fund (twice), CCF-Ant Research Fund, CCF-BaiChuan-Ebtech Foundation Model Fund, Ant Group Research Fund, Tencent Focused Research Fund, Criteo Faculty Research Award, Bytedance Research Collaboration Program, and nomination for Joint AAAI/ACM SIGAI Doctoral Dissertation Award. 
    He also co-organizes DRL4KDD and DRL4IR workshops at KDD'19, WWW'21, SIGIR'20/21/22 and CIKM'23. 

    \item \textbf{Wayne Xin Zhao}\\
    \underline{Email}: batmanfly@gmail.com\\
    \underline{Affiliation}: Renmin University of China\\
    \underline{Biography}:
    Dr. (Wayne) Xin Zhao is currently a professor at Renmin University of China. He obtained the doctoral degree from Peking University in July 2014. He has broad research interest in the fields of information retrieval and natural language processing, with 100+ published papers at top-tier conferences/journals and 10000+ academic citations from Google Scholar. 
    He received ECIR 2021 Test of Time Award, RecSys 2022 Best Student Paper Runner-up, CIKM 2022 Best Resource Paper Runner-up, and other awards. 

    \item \textbf{Yang Song}\\
    \underline{Email}: yangsong@kuaishou.com\\
    \underline{Affiliation}: Kuaishou Technology\\
    \underline{Biography}: Dr. Yang Song is currently the Head of Recommendation at Kwai, overseeing both core modeling and data mining teams. He has published over 70 papers in conferences and journals. He has served as PC\&Area Chairs in Recsys, WSDM, TheWebConf(WWW), IEEE Big Data etc.
    
    \item \textbf{Xiangnan He}\\
    \underline{Email}: xiangnanhe@gmail.com\\
    \underline{Affiliation}: University of Science and Technology of China\\
    \underline{Biography}: Dr. Xiangnan He is a professor at the University of Science and Technology of China (USTC). 
    He has over 100 publications appeared in top conferences such as SIGIR, WWW, and KDD, and journals including TKDE, TOIS, and TNNLS. His work on recommender system has received the Best Paper Award Honourable Mention in SIGIR (2021, 2016) and WWW (2018). He has rich experience in organizing workshops and tutorials at SIGIR'18, WSDM'19\&20, WWW'21\&22, and RecSys'21. 


    \end{itemize}

\bibliographystyle{ACM-Reference-Format}
\bibliography{bibtex}

\end{document}